\documentclass[%
mathleft,%
]{an}
\usepackage{graphicx}
\usepackage{times}

\begin{document}

\Pagespan{589}{}
\Yearpublication{2010}%
\Yearsubmission{2010}%
\Month{1}%
\Volume{331}%
\Issue{6}%
\DOI{This.is/not.aDOI}%

\title{IR Spectroscopy of COmosphere Dynamics with the CO First Overtone Band}
\author{T.A. Schad\inst{1,2}\fnmsep\thanks{Corresponding author:\email{schad@noao.edu}\newline}
\and  M.J. Penn\inst{1}}
\titlerunning{Infrared Spectroscopy of COmosphere Dynamics}
\authorrunning{T.A. Schad \& M.J. Penn}
\institute{National Solar Observatory, 950 N Cherry Ave, Tucson, Arizona 85719 USA
\and Lunar and Planetary Laboratory, University of Arizona, Tucson, Arizona 85721 USA}
\received{2010 Jan 19}
\accepted{2010 Mar 29}
\publonline{2010 Jun 17}


\abstract{We discuss observations of the weak first overtone ($\Delta\nu=2$) CO absorption band near 2300 nm with the U.S. National Solar Observatory Array Camera (NAC), a modern mid-infrared detector.  This molecular band provides a thermal diagnostic that forms lower in the atmosphere than the stronger fundamental band near 4600 nm.  The observed center-to-limb increase in CO line width qualitatively agrees with the proposed higher temperature shocks or faster plasma motions higher in the COmosphere.  The spatial extent of chromospheric shock waves is currently at or below the diffraction limit of the available C0 lines at existing telescopes. Five minute period oscillations in line strength and measured Doppler shifts are consistent with the p-mode excitation of the photospheric gas.  We also show recent efforts at direct imaging at 4600 nm.  We stress that future large-aperture solar telescopes must be teamed with improved, dynamic mid-infrared instruments, like the NAC, to capitalize on the features that motivate such facilities.}
\maketitle

\section{Introduction}

The presence of cool molecular CO at temperatures below the previously established temperature minimum has challenged our understanding of the basic thermal structure of the solar atmosphere and undermined our spectral diagnostics of cool stellar atmospheres (Ayres 2002, and references therein).  Recent simulations of the lower solar atmosphere show the CO forming region, dubbed the ``COmosphere'' by Wiedemann et al. (1994), to be highly dynamic; fine scale structure (scales as small as 0.1$''$) dominates the evolution and energy exchange within the inhomogeneous COmosphere architecture (Wedemeyer-B\"{o}hm et al. 2005).  The ability of these simulations to interpret the observed, spatially and temporally averaged spectra is limited by the accuracy to which nonequilibrium CO chemistry can be incorporated (Asensio Ramos et al. 2003).  For now, the nature of these CO "clouds" remains elusive as not even the spatial extent of these features has been clearly classified.

Diagnostic avenues towards probing the thermal structure of the solar atmosphere have informed the technical requirements of the next generation of solar telescopes.  New 4-meter class all-reflective facilities will finally allow access to the infrared spectral regime at a resolution more comparable to length scales of interest.  In particular, they will be able to exploit the large number of molecular CO lines near 4600 nm ($\Delta\nu=1$) and 2300 nm ($\Delta\nu=2$), which sense quite well the cool component of the mid to upper photosphere (Ayres 2002).  In addition, a number of mid-IR lines with propitious magnetic sensitivity will also be available for more accurate magnetic field determination.

Advancing mid-infrared instrumentation and observing techniques is a clear, but difficult, first step in realizing the full potential of the next class of large aperture solar telescopes.  In the continued observational efforts to resolve the puzzles of the COmosphere, only recently has the available instrumentation allowed investigation of p-mode and granulation dynamics in the CO molecular bands with a reasonable field-of-view (Ayres et al. 2008).  Here we present such an observation using scanned slit-spectroscopy of the first overtone band of CO near 2300 nm to discuss p-mode gas excitation.  We aim to demonstrate the dynamic range of the U.S. National Solar Observatory Array Camera (NAC), which represents a step forward for mid-infrared detection.  We further show significant advances in direct imaging near 4600 nm.  The next generation of IR detectors together with large-aperture solar telescopes should grant insight into the solar atmospheric thermal structure as well as the many other questions that the infrared regime can address.

\section{Observations and data analysis}

\subsection{NSO Array Camera (NAC) observations}

Temporally and spatially resolved spectroscopy as well as direct imaging in the mid-IR are made possible with the NAC at the 1.52 meter main telescope of the McMath-Pierce facility (McM/P) on Kitt Peak, AZ, USA.  The camera's 1024$^{\rm 2}$ InSb Aladdin III Detector offers high density sampling (0.018$''$ pix$^{\rm -1}$) and is sensitive at 1--5 microns. Its helium-cooled focal plane (stable at $30\pm0.01$K) and internal spectral filters ensure a low dark current and a low read out noise allowing unprecedented quality of daytime mid-infrared observations.  High cadence operation (up to 11 Hz) and short exposure capability (down to 25$\mu$s) make this instrument especially useful for observations of dynamical processes.  As an additional advantage, the NAC system can achieve near diffraction-limited observations with the Infrared Adaptive Optics (IRAO) system when a high-contrast target is available.  Although imaging in the mid-IR regime has been historically challenging, the NAC is highly-suited and primed to serve a key pathfinding role for future IR instruments yet to be incorporated in new large aperture solar telescopes.

\begin{figure}
\includegraphics[width=0.45\textwidth,height=50mm]{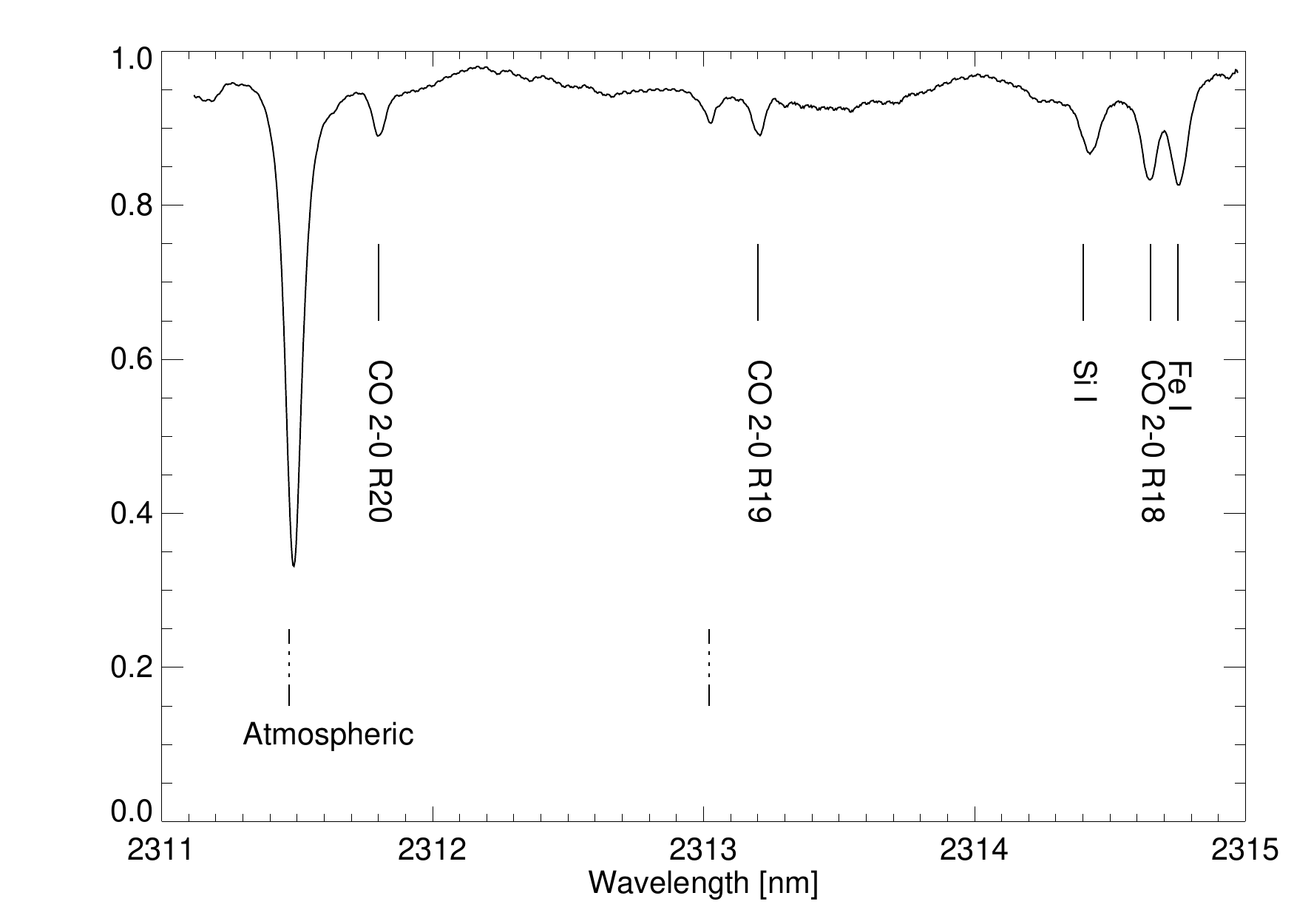}
\caption{NAC reference spectrum near 2313 nm.}
\label{fig:spectrum}
\end{figure}

Here we employ the NAC instrument in tandem with the 13.7 meter Vertical Spectrograph at the McM/P to obtain spectroheliograms near 2312 nm on 2009 February 6 for a large portion of the solar disk, which exhibited no significant activity on this date.  The McM/P prime focal image is reimaged with flat mirrors and a 4 inch lens in order to fit the full solar diameter on the spectrograph slit. This setup reduces the diffraction-limited spatial resolution at 2300 nm to 5.8$''$, which is still smaller than the $\sim$7$''$-8$''$ p-mode ``wavepackets'' (Deubner 1969). Full scans with 500 5$''$ steps were taken at a 75 sec cadence from 15:34 to 17:39 UT.  The NAC instrument was used to record spectra in the 2311--2315 nm range, which includes three first overtone CO absorption lines (see Fig.~\ref{fig:spectrum}). Si I and Fe I lines are also present.  The raw observations have a spatial sampling of 1.5$''$ along the slit and 5$''$ in the scanning direction (i.e. from heliographic south to north in fly-back mode).  Temporal cadence may be  optimized with other scanning modes.

\subsection{Analysis}

Telescope pointing error induces solar image drift perpendicular to the spectrograph slit during our observations.  As the drift is slow ($\approx$ 0.36$''$ min$^{\rm -1}$), we realign each individual scan to sub-pixel accuracy ($<$ 1.5") according to shifts determined by cross-correlating the continuum limb position in each scan.  The spectra are then analyzed according to the intended use.  First, to derive the center-to-limb behavior, the optimal signal-to-noise (S/N) is obtained by coadding all spatially-local spectra at the original $5''\times1.5''$ sampling resolution. Each spectrum, however, is shifted prior to coadding to correct for its local Doppler shift.  The resulting S/N of the coadded spectra is 1400. The CO absorption lines at 2311.8 and 2313.2 nm are then fit with a linear combination of a gaussian function with a second-order polynomial, which estimates the background contribution.  

To study the observed spectra as a time series, each raw isotemporal spectrum is smoothed spatially and spectrally (5$\times$5 pixel boxcar smoothing), resulting in a S/N of 1200.  The Doppler shift ($\Delta\lambda$) and residual intensity (I$_{\rm res}$) of each CO absorption line are then measured, and spatial maps of $\Delta\lambda$ and I$_{\rm res}$ with $\sim8''\times6''$ spatial resolution are created.  A common time is established for each individual map using linear interpolation in time.

\section{Results}

\subsection{Center-to-limb behavior}


The observed center-to-limb behavior of the CO 2-0 R20 line equivalent width ($\Delta W / \Delta\mu =-1.2$ m\AA) compares favorably with that shown by Ayres (1978).  Increases in equivalent width towards the limb suggest unresolved velocity and/or temperature fluctuations at greater heights.  This has required an angle-dependent microturbulence parameter to be incorporated within atmospheric models to fit the observed behavior (Ayres 1978). Future large aperture solar telescopes may better address these fluctuations.  The observed center-to-limb line depth variation ($\Delta{\rm dep}/\Delta\mu=0.052$) compared to the half-width of the line depth spatial histogram ($\sim0.006$) shows behavior similar to the weaker fundamental absorption lines thought to form in the middle photosphere (Ayres \& Rabin 1996).

\subsection{Time series analysis of oscillations}

\begin{figure}
\begin{center}
\includegraphics[width=0.45\textwidth,height=50mm]{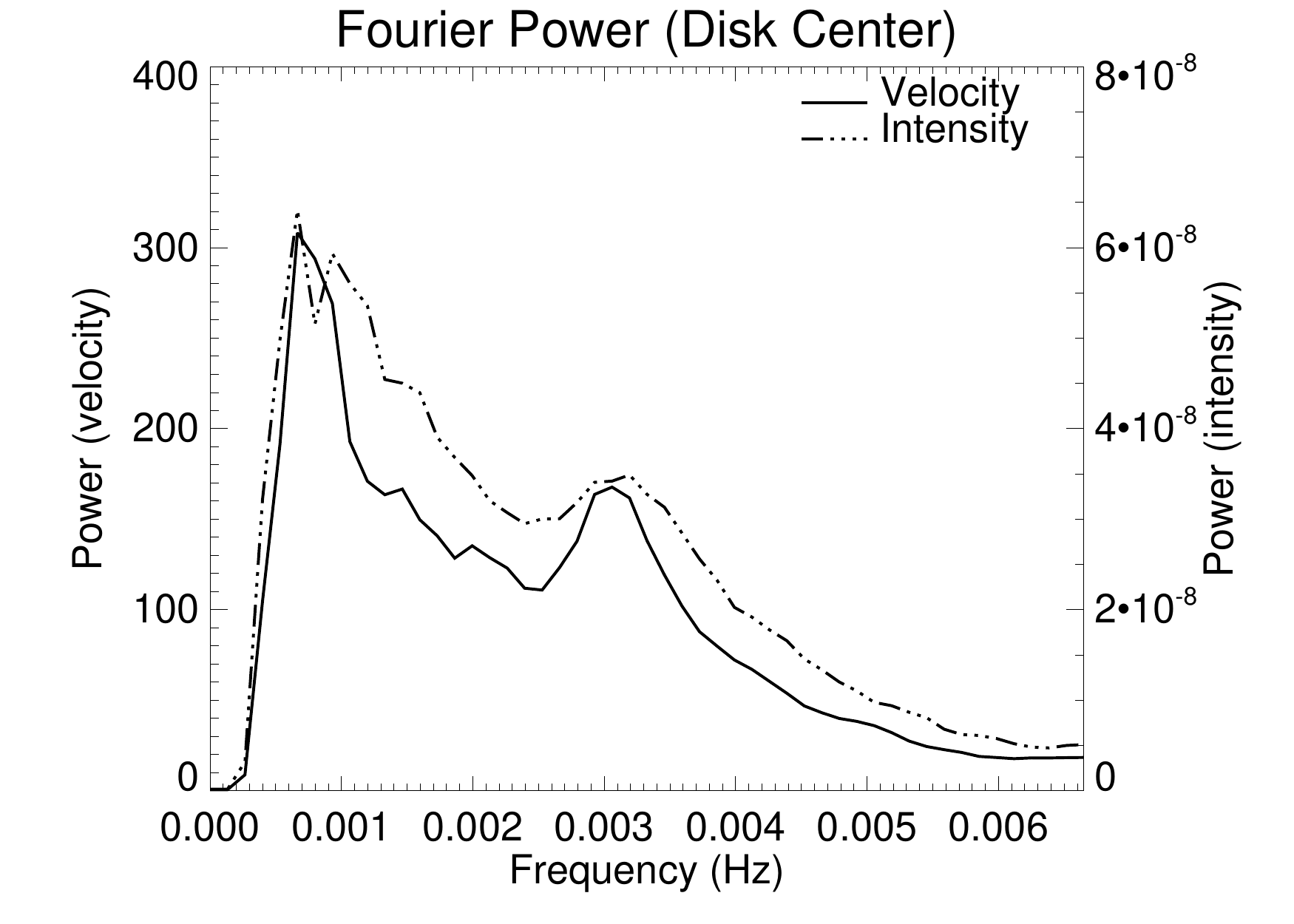}
\includegraphics[width=0.45\textwidth,height=50mm]{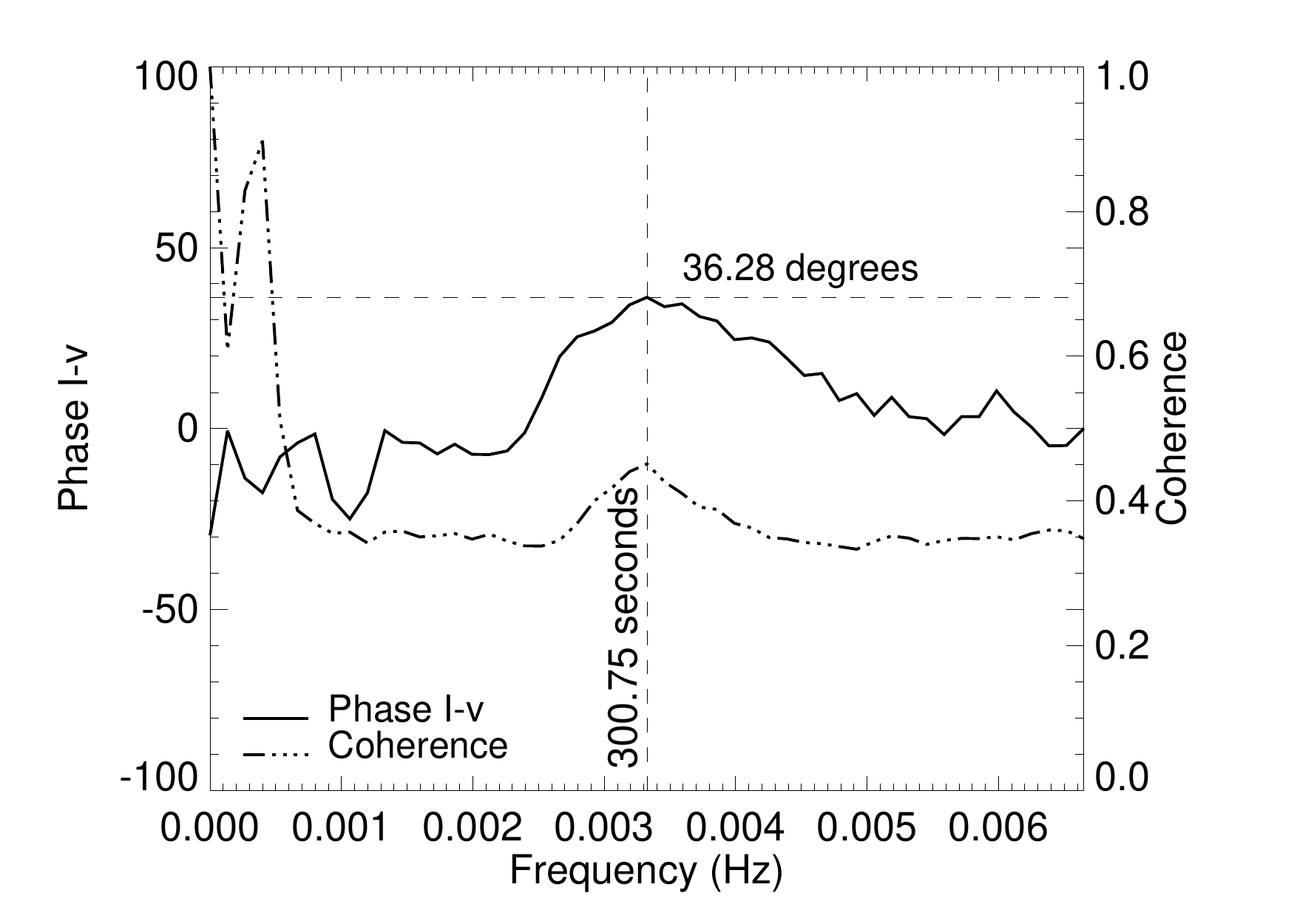}
\end{center}
\caption{Average Fourier power spectrum (top) of CO residual intensity and Doppler shift measurements near disk center for the absorption lines at 2311.8 and 2313.2 nm. Notice strong peak at 3.3 mHz. I-v phase difference and coherence spectra (bottom).}
\label{fig:power_spec}
\end{figure}

Power spectral analysis of temporal residual intensity and Doppler shift variations in the two observed unblended CO absorption lines reveals prominent five minute period (3.3 mHz) oscillations observed near disk center consistent with p-modes (see Fig.~\ref{fig:power_spec}).  The power spectrum shown results from the average power spectrum of each quantity, each line, and each spatial pixel in a 150$''\times$250$''$ region near disk center to reduce noise.  Each time series is first apodized and a polynomial fit to the background trend is subtracted.  Power is insignificant below the frequency resolution (0.8 mHz) and exhibits a noise background above the minimum frequency resolved. For observations near the limb (not shown), the Fourier power of measured Doppler shifts nearly vanishes, as expected.  Power at 3.3 mHz is still apparent but weakened for intensity measurements.  With improvements in noise reduction and resolution, k-$\nu$ diagnostic diagrams should be attainable.  CO line Dopplergrams (with solar rotation subtracted) do show large scale structures with horizontal flow velocities consistent with supergranules.

Significant cross power exists between the intensity and velocity fluctuations at the five minute period.  The temporal coherence between the two variables is lower than the typical 0.6 cutoff, but a local peak in coherence is evident at a five minute period.  At this frequency the phase lag between the intensity maximum and Doppler redshift maximum is $\sim36\degr$, similar to the weaker fundamental CO lines (Ayres \& Brault 1990) and consistent with the near adiabatic response of the mid-photospheric gas to evanescent waves (Noyes \& Hall 1972).

\section{Direct mid-infrared imaging}

Despite complications invoked by the large solar infrared flux, mid-infrared direct imaging shows considerable promise and has been in development at the McM/P for nearly two decades (Livingston et al. 1992; Ayres et al. 2008).  Infrared direct imaging capabilities are improving with the deployment of the NAC instrument.  Recent imaging of K-band (2200 nm) continuum granulation with the NAC has verified that spatial resolution near the diffraction limit is possible at the McM/P at longer wavelengths (Penn 2008), which boast a decrease in seeing and stray light compared to the visible (Boyd 1978).  The McM/P resolution (0.76$''$ at 4600 nm), however, is still incapable of resolving the spatial scales of interest.  Recently, direct imaging at 4600 nm has been undertaken using the NAC.  NOAA active region 11027 was imaged in 2009 September under good observational conditions without adaptive optics (see Fig. 3).  The 15 nm wide spectral filter requires neutral filtering to ensure the array does not saturate.  Beyond some preliminary flat and dark correction, no post processing has been implemented.

\begin{figure}
\begin{center}
\includegraphics[width=0.45\textwidth,height=0.45\textwidth]{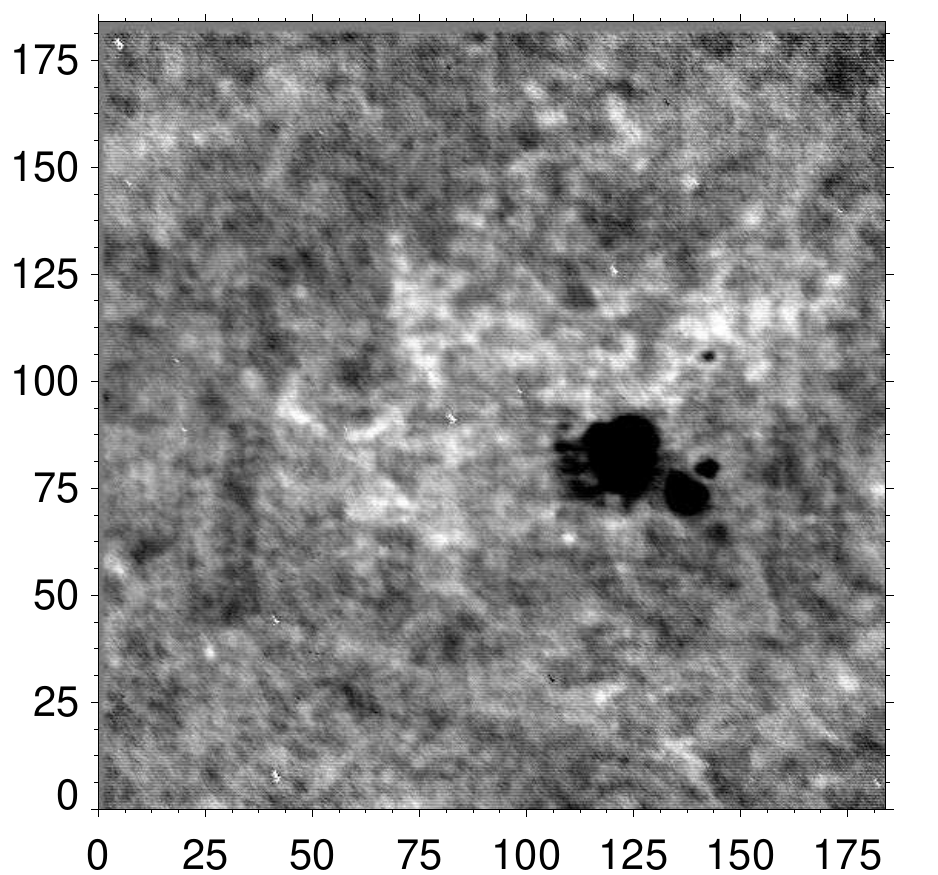}
\end{center}
\caption{Filtered image of NOAA 11027 at 4600 nm acquired by the NSO Array Camera (NAC) at the McMath-Pierce Facility.  Solar north is up, and tickmarks correspond to seconds of arc.}
\label{fig:co_image}
\end{figure}

\section{Concluding remarks}

The NAC proves an apt detector to explore mid-IR dynamics; however, the relevant scales are still unattainable with existing facilities.  The dawn of large aperture solar telescopes offers great promise to advance daytime infrared astronomy, and with it, our understanding of the solar atmosphere.  We encourage the development of future IR instruments to take full advantage of this spectral regime.

\acknowledgements
We thank the U.S. National Science Foundation for supporting Tom Schad's participation in the 1st EAST-ATST workshop as a graduate student. This work was supported by the U.S. National Solar Observatory.


\begin{thebibliography}{}
\bibitem[\protect\citeauthoryear{Asensio Ramos et al.}{2003}]{asensio03}Asensio Ramos, A., Trujillo Bueno, J., Carlsson, M., Cernicharo, J.: 2003, \apj 588, L61
\bibitem[\protect\citeauthoryear{Ayres}{1978}]{ayres78}Ayres, T.R.: 1978, \apj 225, 665
\bibitem[\protect\citeauthoryear{Ayres \& Brault}{1990}]{ayres90}Ayres, T.R., Brault, J.W.: 1990, \apj 363, 705 
\bibitem[\protect\citeauthoryear{Ayres \& Rabin}{1996}]{ayres96}Ayres, T.R., Rabin, D.: 1996, \apj 460, 1042 
\bibitem[\protect\citeauthoryear{Ayres}{2002}]{ayres02}Ayres, T.R.: 2002, \apj 575, 1104
\bibitem[\protect\citeauthoryear{Ayres et al.}{2008}]{ayres08}Ayres, T., Penn, M., Plymate, C., Keller, C.: 2008, 12th European Solar Physics Meeting, Freiburg, Germany, p.2.74, 12, 2 
\bibitem[\protect\citeauthoryear{Boyd}{1978}]{boyd78}Boyd, R.W.: 1978, JOSA 68, 877 
\bibitem[\protect\citeauthoryear{Deubner}{1969}]{deubner69}Deubner, F.-L.: 1969, SoPH 9, 343
\bibitem[\protect\citeauthoryear{Livingston et al.}{1992}]{livingston92}Livingston, W., Kopp, G., Gezari, D.: 1992, \baas, 24, 1252 
\bibitem[\protect\citeauthoryear{Noyes \& Hall}{1972}]{noyes72}Noyes, R.W., Hall, D.N.: 1972, \apj 176, L89
\bibitem[\protect\citeauthoryear{Penn}{2008}]{penn08}Penn, M.J.: 2008, American Geophysical Union Spring Meeting, abstract \#SP41B-01
\bibitem[\protect\citeauthoryear{Wedemeyer-B\"{o}hm et al.}{2005}]{wedemeyer05}Wedemeyer-B\"{o}hm, S., Kamp, I., Bruls, J., Freytag, B.: 2005, A\&A 438, 1043 
\bibitem[\protect\citeauthoryear{Wiedemann et al.}{1994}]{wiedemann94} Wiedemann, G.R., Ayres, T.R., Jennings, D.E., Saar, S.H.: 1994, \apj 423, 806
\end{thebibliography}
\end{document}